\documentclass[sigconf]{acmart}
\usepackage{booktabs}
\usepackage{multirow}
\usepackage{graphicx}
\usepackage{subfigure}
\usepackage{xcolor}
\usepackage{adjustbox}

\makeatletter
\g@addto@macro\normalsize{%
  \abovedisplayskip 6pt plus 2pt minus 2pt%
  \belowdisplayskip \abovedisplayskip
  \abovedisplayshortskip 6pt plus2pt  minus2pt%
  \belowdisplayshortskip 6pt plus2pt minus2pt%
}
\makeatother

\copyrightyear{2020}
\acmYear{2020}
\setcopyright{acmcopyright}
\acmConference[WSDM '20]{The Thirteenth ACM International Conference on Web Search and Data Mining}{February 3--7, 2020}{Houston, TX, USA}
\acmBooktitle{The Thirteenth ACM International Conference on Web Search and Data Mining (WSDM '20), February 3--7, 2020, Houston, TX, USA}
\acmPrice{15.00}
\acmDOI{10.1145/3336191.3371790}
\acmISBN{978-1-4503-6822-3/20/02}

\begin{document}

\title{Distilling Structured Knowledge into Embeddings for Explainable and Accurate Recommendation}


\author{Yuan Zhang}
\affiliation{%
\institution{Key Laboratory of Machine Perception (MOE)}
  \institution{Department of Machine Intelligence, Peking University}
  }
\email{yuan.z@pku.edu.cn}

\author{Xiaoran Xu}
\affiliation{%
  \institution{Hulu LLC}
  }
\email{xiaoran.xu@hulu.com }

\author{Hanning Zhou}
\affiliation{%
  \institution{Facebook}
  }
\email{ericzhouh@gmail.com}

\author{Yan Zhang}
\affiliation{%
 \institution{Key Laboratory of Machine Perception (MOE)}
  \institution{Department of Machine Intelligence, Peking University}
  }
\email{zhyzhy001@pku.edu.cn}

\renewcommand{\shortauthors}{}

\begin{abstract}
Recently, the embedding-based recommendation models (e.g., matrix factorization and deep models) have been prevalent in both academia and industry due to their effectiveness and flexibility.
However, they also have such intrinsic limitations as lacking explainability and suffering from data sparsity.
In this paper, we propose an end-to-end joint learning framework to get around these limitations without introducing any extra overhead by distilling structured knowledge from a differentiable path-based recommendation model.
Through extensive experiments, we show that our proposed framework can achieve state-of-the-art recommendation performance and meanwhile provide interpretable recommendation reasons.
\end{abstract}

\fancyhead{}

\maketitle
\section{Introduction}
\label{intro}
Extensive research efforts have been dedicated to recommender systems in the past decades. 
Starting from the winning solutions to the Netflix prize challenge, embedding-based recommendation models have attracted most of the attention, in which users and items are represented by latent vectors, such as the matrix factorization (MF) models \cite{koren2009matrix}, factorization machines (FM) \cite{rendle2010factorization}, and their recent neural extensions \cite{covington2016deep, he2017neural, he2017neuralfac}. Their popularity is largely attributed to their superior performance, and also to their flexibility in incorporating both collaborative signals and content-related signals (e.g., \cite{he2016vista, zhang2016collaborative, zhang2017joint, wang2018dkn}).
Nowadays, however, when not merely practitioners but also users are increasingly interested in the underlying reasons for each recommendation \cite{zhang2014explicit, catherine2017explainable}, their explainability becomes a big issue. It is hard for us to know which factor contributes the most and why it leads to the final recommendations outside the black box. 
Furthermore, those embedding-based models suffer from the data sparsity problem and are prone to over-fitting, while cold-start users and long-tail items are inevitable challenges in real-world application scenarios. This issue becomes more important as the model capacity grows in the deep learning era.

Another line of research exploits path-based models \cite{yu2014personalized, shi2015semantic, catherine2016personalized, shi2018heterogeneous} to integrate different recommendation signals, e.g., meta-paths \cite{sun2011pathsim} over heterogeneous information networks (HINs) as shown in Table \ref{tab:path}. These models are tempting because the paths are human-interpretable features and, meanwhile, robust to sparse data with the help of graph structures. Nonetheless, traditional pure path-based models lack sufficient expressive power to obtain comparable performance to the state-of-the-art embedding-based approaches.
Therefore, most existing work integrates embedding-based methods \textit{within} their recommendation pipelines.
However, the introduction of embeddings into path-based models might hurt both their explainability and generalization capability by potentially inheriting the limitations of embedding-based models. In addition, these approaches are generally hard to deploy in a real-time production setting due to their inefficiency.

In this paper, we propose an end-to-end joint learning framework to complement the strengths and weaknesses of the two approaches mentioned above.
As shown in Figure \ref{fig:illustration}, given an embedding-based model within the black box, we jointly train that model with a differentiable pure path-based model. 
Different from regular joint training approaches with shared parameters, our framework only couples these two models in the label space by a mutual regularization term in the objective function.
From one direction, the embedding-based model is \textit{regularized} by the graph-structured knowledge encoded in the path-based model to avoid poorly generalizable local minima and thus make more accurate recommendations. 
From the other direction, we optimize the learnable paths to mimic the given embedding-based model in order to obtain its \textit{interpretation}. 
Besides, no extra computational burden is introduced for either model at test time.

From a higher perspective, the proposed approach can be regarded as \textit{knowledge distillation} proposed by Hinton et al. \cite{hinton2015distilling} with the difference that the teacher model and the student model are learning from each other simultaneously. In other words, the embedding-based model as a student is allowed to distill the structured knowledge from the path-based model in addition to imitation learning from concrete training labels. At the same time, the path-based model as a teacher can also enhance its knowledge encoded in the reasoning paths by synthesizing predictions made by the embedding-based model.

From a more technical perspective, our proposed approach can also be seen as a label propagation method to address missing data problem in recommender systems \cite{ma2007effective, lim2015top}. More specifically, given a user, not all unobserved items are irrelevant to her, but the one-hot labels treat them equally as negative examples. On the contrary, the path-based model can \textit{learn to propagate} sparse labels into unobserved items, where relevant unobserved items are likely to be given relatively higher probability scores to irrelevant ones. Learning from the augmented labels overcomes the data sparsity issue, enables efficient structure learning and hence improves recommendation accuracy.

Finally, we conduct extensive experiments across various publicly available datasets in multiple domains. The experimental results show that the proposed approach achieves state-of-the-art performance. Ablation study further demonstrates that the joint learning procedure leads to substantial improvement over alternative approaches in terms of both accuracy and explainability. Meanwhile, we demonstrate the ability of the proposed framework to generate interpretable recommendations through case studies.

\begin{table}[t]
  \caption{Example meta-paths in HINs for recommendation.}
  \centering
\begin{adjustbox}{max width=\linewidth}
    \begin{tabular}{cc}
    \toprule
    Meta-path & Recommendation model \\
    \midrule
    user $\rightarrow$ item $\rightarrow$ user $\rightarrow$ item & collaborative recommendation \\
    user $\rightarrow$ user$\rightarrow$  item & social recommendation \\
    user $\rightarrow$ item$\rightarrow$  category$\rightarrow$  item & content-based recommendation \\
    \bottomrule
    \end{tabular}%
\end{adjustbox}    
  \label{tab:path}%
\end{table}%

\section{Methodology}
\subsection{Problem Formulation and Preliminaries}
\label{prelim}
In this paper, we focus on the top-N item recommendation task with implicit feedback, that is, only implicit user feedback such as clicks, watch history, purchases\footnote{We sometimes use the word \textit{click} to refer to all these types of interactions in general.} are available in contrast to explicit rating information. Formally, let $U$ be the set of users and $I$ the set of items. We observe user-item interactions $D = \bigcup_{u \in U} \{(u, i) | i \in I_u\}$, where $I_u \subseteq I$ is the set of items user $u$ has clicked. In some cases, we are also given $L_U$ and $L_I$ types of attributes of users and items as metadata (e.g., social relations, genres, actors and directors) denoted as $\{C_u^{(l)} \subseteq C^{(l)}_U |l = \{1, ..., L_U\}, u \in U\}$ and $\{C_i^{(l)} \subseteq C^{(l)}_I|l =\{1, ..., L_I\}, i \in I\}$, where $C^{(l)}_U$ and $C^{(l)}_I$ are the vocabularies of the $l$-th type of attributes for users and items, respectively. Our task is to recommend each user $u$ a list of $N$ relevant items from $I \setminus I_u$. 

Following \cite{liang2018variational}, we use binary encoding $\mathbf{x_u} \in \{0,1\}^{|I|}$ to denote items that user $u$ has clicked. The recommendation algorithm $f(\mathbf{x_u})$ produces a probability distribution $\mathbf{z_u}$ over $I$ to capture the relevance of items to user $u$. At test time, for each user $u$, we simply rank items in $I \setminus I_u$ with their relevance scores given in $\mathbf{z_u}$ for top-N recommendation. Note that $\mathbf{x_u}$ can also be regarded as an unnormalized empirical distribution (we also denote its normalized form as $\tilde{\mathbf{x}}_{\mathbf{u}}$) with the difference that $\mathbf{x_u}$ is much sparser than $\mathbf{z_u}$.

Although our proposed framework is not limited to any specific embedding-based recommendation model $f(\cdot)$, for ease of presentation, we introduce a simple one as an example throughout this paper. Specifically, we represent each user $u \in U$ with 
\begin{equation}
\mathbf{v_u} = \tanh(\mathbf{r_u} + \frac{1}{|I_u|}\sum_{i\in I_u} \mathbf{r_i} + \sum_{l=1}^{L_U}\frac{1}{|C_u^{(l)}|}\sum_{c \in C_u^{(l)}} \mathbf{r_c}),
\end{equation}
where $\mathbf{r_u}, \mathbf{r_i}, \mathbf{r_c} \in \mathbb{R}^{d}$ are $d$-dimensional embedding vectors for user $u$, item $i$ and metadata attribute $c$, respectively.
Similarly, we represent each item $i \in I$ with
\begin{equation}
\mathbf{v_i} = \tanh (\mathbf{r_i} + \sum_{l=1}^{L_I}\frac{1}{|C_i^{(l)}|}\sum_{c \in C_i^{(l)}} \mathbf{r_c}). 
\end{equation}
Then, similar to GMF \cite{he2017neural}, we feed the element-wise product of user representation $\mathbf{v_u}$ and item representation $\mathbf{v_i}$ into a linear layer to get the relevance score for each user-item pair $(u, i)$ with weight $\mathbf{w} \in \mathbb{R}^{d}$
\begin{equation}
r_{u,i} = \mathbf{w}^{T} \cdot(\mathbf{v_u} \odot \mathbf{v_i}).
\end{equation}
Finally, we use softmax to obtain the user preference distribution $\mathbf{z_u}$ for each u,
\begin{equation}
\mathbf{z_u} = softmax([r_{u,i_1}, r_{u,i_2}, ..., r_{u, i_N}]).
\end{equation}

\begin{figure}[t]
    \centering
	\includegraphics[width=8cm]{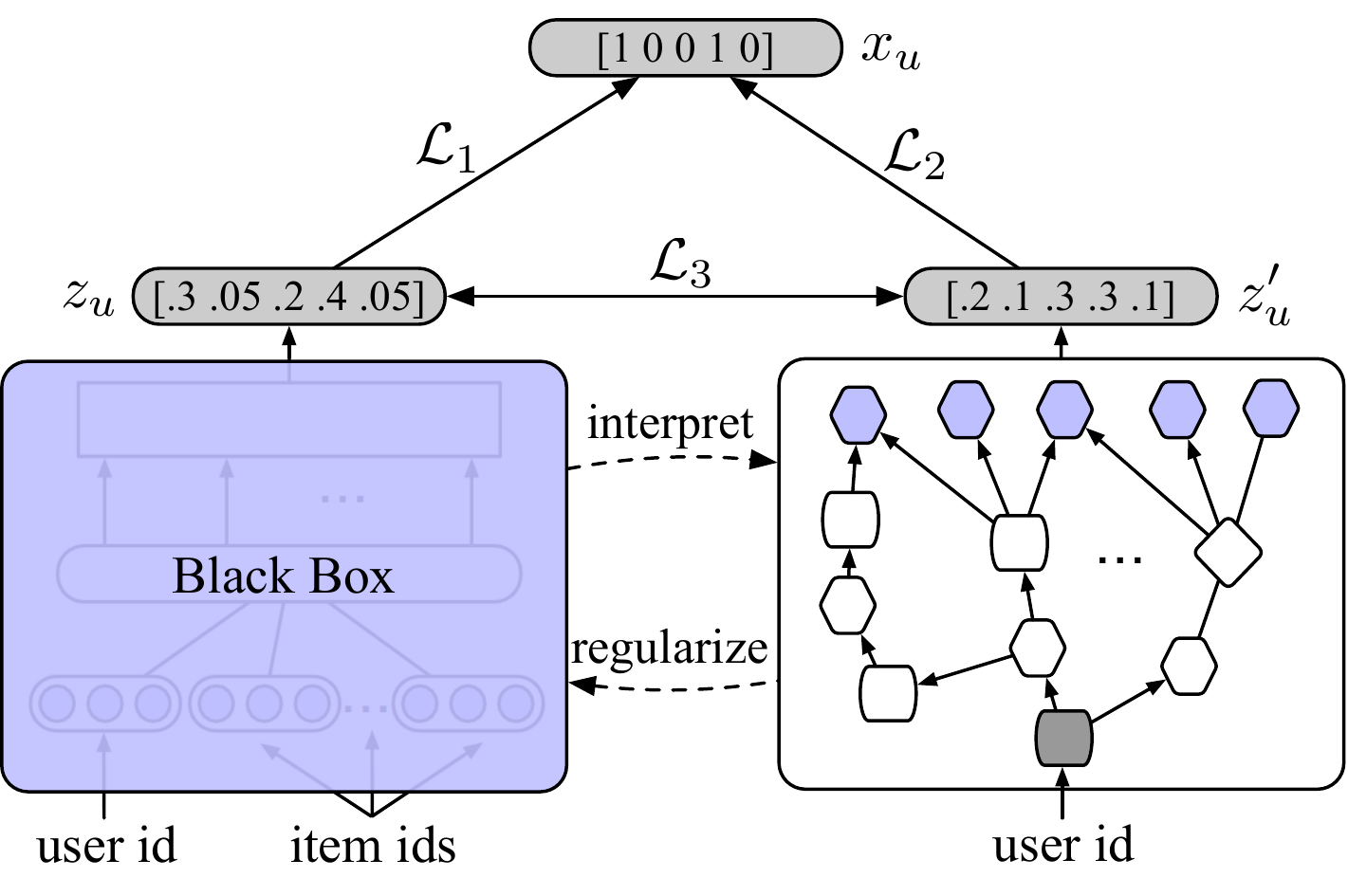}
	\caption{An illustration of the proposed framework.}
	\label{fig:illustration}
\end{figure}

To train this embedding-based model, we optimize the cross-entropy between the normalized ground-truth label $\tilde{\mathbf{x}}_{\mathbf{u}}$ and the output distribution $\mathbf{z_u}$ for each user $u$. 
\begin{equation}
\label{l1}
\min \mathcal L_1 = \sum_{u\in U}\sum_{i\in I} - \tilde{x}_{ui} \log z_{ui}.
\end{equation}
However, this is troublesome for the following two reasons. First, the sparse ground-truth labels $\{\mathbf{x_u}\}$ treat all the unobserved (i.e., not clicked) items equally, while those unobserved items are by no means all irrelevant to that user \cite{hsieh2017collaborative} (otherwise, there is no point making recommendations). This makes it difficult, at least inefficient, for the embedding-based model to learn the underlying structure in data. Even worse, the model is prone to over-fitting, especially for sparse data. Suppose that the capacity of the model were sufficiently large, we could overfit the objective function by putting large probability mass on observed items while putting small yet uninformative (or even arbitrary in the extreme case) probability mass on unobserved ones. This issue is sometimes referred to as the missing data problem \cite{ma2007effective,lim2015top} or the one-class problem \cite{pan2008one} in recommender systems.

Although these limitations seem to be intrinsic to the recommendation task, there are ways to mitigate them such as regularization with weight decay and dropouts. In addition, Pan et al. also propose to use handcrafted weighting schemes and negative sampling strategies \cite{pan2008one}. The proposed framework in this paper provides a more principled and effective approach utilizing graph structures to overcome this issue.

\subsection{Model Framework}
\subsubsection{Differentiable Path-based Model}
\label{diff_path}
In our framework, we introduce a differentiable path-based model to help with interpretation and structure learning for embedding-based models. Different from most existing work, e.g., \cite{yu2014personalized, catherine2016personalized, shi2018heterogeneous}, we decouple our path-based model from embeddings. Instead, inspired by Cohen \cite{cohen2016tensorlog}, we associate each edge with a learnable transition probability to enhance its expressive power while still preserving interpretability. 

More specifically, we first construct a heterogeneous information network (HIN) by regarding users, items and available metadata attributes as different types of nodes and regarding their interactions as edges. Then we define some useful meta-paths \cite{sun2011pathsim} (paths over node types) $\mathcal P=\{P_1, ..., P_K\}$  to find relevant items for users. For example, ``$P_1: user \xrightarrow{\text{clicked}} item \xrightarrow{\text{clicked}^{-1}} user \xrightarrow{\text{clicked}} item$'' is a meta-path for the collaborative recommendation and ``$P_2: user \xrightarrow{\text{clicked}} item \xrightarrow{\text{contains}^{-1}} category \xrightarrow{\text{contains}} item$'' is a meta-path for the content-based recommendation. Every single step within a meta-path can be represented by a transition matrix between two types of nodes. That is, for each edge ``$A \xrightarrow{R} B$'', we denote $M_{A, B}^{R}$ as the transition matrix between nodes with type $A$ and type $B$ (the superscript $R$ is omitted when it does not cause ambiguity) and for each $i \in A$ and $j \in B$
\begin{equation}
M_{A,B}^{R}[i, j] =
\begin{cases}
p_{ij} & (i, j) \in R\\
0 & \text{otherwise},\\
\end{cases}
\end{equation}
where $\{p_{ij} | (i,j)\in R\}$ are learnable parameters attached to each edge. 
Similar to \cite{cohen2016tensorlog}, we do not impose constraints on parameters to ensure the transition matrix is indeed a probability matrix (a.k.a. stochastic matrix). Instead, we only require them to be nonnegative weights by parameterizing $p_{ij} := \sigma(\rho_{ij})$, where $\sigma$ can be \textit{Relu} or \textit{Softplus}. At the end, we normalize the final diffused scores into a distribution as we will show in Eq. (\ref{eq:path}).
For the inverse relation $R^{-1}$, we can use either a separate matrix $M_{A, B}^{R^{-1}}$ or simply its transpose. We choose the latter in our work.

Therefore, the transition matrix $\Pi_P$ of a random walker reaching item nodes from user nodes through a meta-path $P \in \mathcal P$ can be computed as the product of the transition matrices along the path, e.g., $\Pi_{P_1} = M_{UI}\cdot M_{UI}^{T}\cdot M_{UI}$ and $\Pi_{P_2} = M_{UI} \cdot M_{CI}^{T} \cdot M_{CI}$ for the previous examples.
We also introduce a probability distribution $\mathbf p = \{p_{P_1}, ..., p_{P_K}\}$ over all meta-paths to account for different contributions of different meta-paths. Finally, the relevance of items to a given user $u$ is modeled as the probability of a random walker walking from that user through all possible meta-paths to targeted items, formally as,
\begin{equation}
\mathbf{z'_u} = \sum_{k=1}^{K} p_{P_k} \cdot \big(\Pi_{P_{k}}[u, :] \big/ Z_{u, k}\big) = \sum_{k=1}^{K} p_{P_k} \cdot \big(\mathbf{e_u} \Pi_{P_{k}} \big/ Z_{u, k}\big),
\label{eq:path}
\end{equation}
where $Z_{u, k} \in \mathbb{R}$ is the normalization factor to ensure $\Pi_{P_{k}}[u, :]$ is a normalized distribution over items and $\mathbf{e_u}$ is the one-hot row vector of user $u$, i.e., $e_u[u] = 1$ and $e_u[u'] = 0$ for $u' \neq u$.

As we can see, the output distributions $\{\mathbf{z'_u}\}$ can be unrolled into a series of matrix operations and thus are differentiable with respect to all the free parameters in this model. It allows us to train the meta-path guided random walk model with gradient-based optimizers by minimizing the cross-entropy between the ground-truth labels $\tilde{\mathbf{x}}_{\mathbf{u}}$ and the model output $\mathbf{z'_u}$,
\begin{equation}
\label{l2}
\min \mathcal L_2 = \sum_{u\in U}\sum_{i\in I} - \tilde{x}_{ui} \log z'_{ui}.
\end{equation}

This model can be seen as a learnable label propagation method that transforms sparse binary labels $\{\mathbf{x_u}\}$ into dense distributions $\{\mathbf{z'_u}\}$.
The items, that are not clicked but connected by some meta-paths over the HIN, now can also be considered somewhat more relevant than random ones. For example, if someone has only watched one movie, the binary label is simply the one-hot encoding of that movie, while, in the more informative dense distribution, appropriate scores can be propagated into the movies with the same genres (through ``$user\rightarrow genre \rightarrow user$'') and the ones that are frequently co-watched by other users (through ``$user \rightarrow (movie \rightarrow user)^n$'') as well. (In fact, the latter example is the main motivation for \textit{collaborative metric learning}\cite{hsieh2017collaborative} where relevance transitivity is attempted to be maintained.) Using them as augmented pseudo-labels can help downstream models with efficient structure learning.

\subsubsection{End-to-end Joint Learning Framework}
To address the limitations of the embedding-based recommendation models, we introduce an end-to-end joint learning framework with the differentiable path-based model as shown in Fig. \ref{fig:illustration}. In addition to minimizing the cross-entropy losses of the embedding-based model and the path-based model separately as in Eq. (\ref{l1}) and (\ref{l2}), we also wish to minimize the distance between these two models measured by the KL-divergence between their output distributions,
\begin{equation}
\label{kl}
\min_{\mathbf{z}, \mathbf{z'}}  \mathcal L_3 = \sum_{u \in U} KL(\mathbf{z'_u}||\mathbf{z_u}) = \sum_{u \in U} \sum_{i \in I} - z'_{ui} \log z_{ui} + z'_{ui} \log z'_{ui}
\end{equation}

Note that we optimize the loss function in Eq. (\ref{kl}) with respect to both the embedding-based model $\mathbf{z}$ and the path-based model $\mathbf{z'}$. By optimizing w.r.t. the former, we are actually minimizing the first term in the RHS of Eq. (\ref{kl}) which coincides with the cross-entropy loss with the output distributions $\{\mathbf{z'_u}\}$ as pseudo-labels. In other words, our approach allows the embedding-based model to learn from two sources, i.e., the observed user-item interactions as in Eq. (\ref{l1}) and the augmented pseudo-labels by meta-path guided random walks. 
At the same time, by optimizing w.r.t. the latter, we are trying to find a projection in the parameter space of our differentiable path-based model so that these two models make the most consistent predictions. In this direction, we can not only obtain an interpretation of the embedding-based model within the black box, but also allow the differentiable path-based model to provide more accurate pseudo-labels in a bootstrapping manner. Overall, this joint learning process regularizes the embedding-based model to search for more generalizable and explainable local minima (see more discussion in Sect. \ref{eval_joint}).

Our proposed approach can be seen as \textit{knowledge distillation} \cite{hinton2015distilling}. The differentiable path-based model plays the role of the teacher in our framework. It synthesizes the training labels to enhance its domain knowledge encoded in meta-paths to provide more informative training signals for the student model. Meanwhile, the embedding-based model as the student ``distills'' the structured knowledge within the teacher model into its embedding parameters. One difference from the original distillation work is that we let the teacher model and the student model learn from each other rather than only in one direction because they both have their own merits as mentioned in Sect. \ref{intro}.
\subsubsection{Model Training}
\label{training}
To sum up the previous sections, the loss function used in the training stage is 
\begin{equation}
	\mathcal L = \alpha \mathcal L_1 + \beta \mathcal L_2 + \mathcal L_3,
\end{equation}
where $\alpha$ and $\beta$ are imitation parameters controlling the balance between imitation learning (from concrete training labels) and knowledge distillation. Motivated by Hu et al. \cite{hu2016harnessing}, we exploit dynamic imitation parameters that change as training goes on. In the beginning, we use relatively large imitation parameters $\alpha$ and $\beta$ since the teacher model is still enhancing its own knowledge (with $\mathcal L_2$) and the training labels are still informative to a randomly initialized embedding-based model to ``grasp some basics'' (with $\mathcal L_1$). As the training process continues, we decrease the imitation parameters to allow the student model to distill structured knowledge into its parameters by interacting with the augmented pseudo-labels generated by the teacher model (with $\mathcal L_3$).

In practice, there are two ways to schedule the imitation parameters. A simple approach is to set the imitation parameters to relatively large values $\eta_0$ in the first $T_0$ iterations and suddenly decrease them to smaller values $\eta_1$ afterward. A rule of thumb would be to select $T_0$ to be the time when the validation loss of the embedding-based model starts to bounce back (meaning the training labels are no longer informative and begin to cause over-fitting as discussed in Sect. \ref{prelim}). Another way is to gradually decrease them with a decaying function, e.g., $\eta(t) = \max(\eta_0 \cdot \lambda^{t}, \eta_1)$ where $\lambda \in (0, 1)$ is a tunable decay factor. We find that the first approach already works surprisingly well enough, while the latter can sometimes bring about marginal improvements.

The proposed joint training framework shares a similar idea to DML (Deep Mutual Learning) \cite{zhang2018deep} recently proposed in computer vision. As discussed in \cite{zhang2018deep}, the mutual learning framework can help to find more robust local minimal by entropy regularization.
Different from DML which jointly trains two neural architectures (eg., Resnet and MobileNet), our approach combines two types of models, i.e., embedding/neural models and path/graph models, with rather different inductive biases \cite{zhang2019neural} to allow them to benefit more from each other.

\subsubsection{Complexity and Scalability}
The major time complexity comes from the path-based model. At first glance, it makes the proposed approach inefficient to compute $\mathbf{z'_u}$ in Eq. (\ref{eq:path}) with extensive matrix multiplications. 
However, we note that transition matrices are usually extremely sparse. Thus, we can unroll $\mathbf{e_u}\Pi_{P_k}$ and compute it with a series of sparse matrix-vector multiplications. Since each sparse matrix-vector multiplication takes linear time complexity w.r.t. the non-zero elements within the sparse matrix, our algorithm scales linearly w.r.t. the number of relations as well.

\subsection{Explainable Recommendation}

We are able to interpret recommendations on the following two levels. 
Firstly, given an item $i$ recommended to user $u$, we can tell which factor contributes the most by computing the contribution of each meta-path $P$ as
\begin{equation}
	w_P(u, i) = \frac{p_{P} \cdot \Pi_P[u, i]}{\sum_{P'\in \mathcal P} p_{P'} \cdot \Pi_{P'}[u, i]}.
\end{equation}
For example, $w_{P_1}(u, i)$ and $w_{P_2}(u, i)$ measures the contribution of collaborative signals and category information, as in the example in Sect. \ref{diff_path}, respectively. This level of interpretation is especially useful when the recommender system exploits various types of metadata. 
After we know the most important factors, we can then apply the \textit{beam search} algorithm to find concrete paths with high probabilities to explain at a more detailed level (see Table~\ref{tab:case}).

\section{Experiments}
\subsection{Overview}
In this section, we empirically study the effectiveness of the proposed framework. Specifically, we are interested in the following research questions:
\begin{itemize}
\item[\textbf{RQ1.}] How does our approach compare to the state-of-the-art baseline models?
\item[\textbf{RQ2.}] Is the proposed approach able to provide interpretable recommendation results?
\item[\textbf{RQ3.}] How does joint learning help with recommendation in terms of explainability and accuracy?
\end{itemize}

\subsection{Experimental Setup}
\subsubsection{Datasets}
Our proposed approach is evaluated on the following public available datasets in various domains.
\begin{itemize}
  \item \textbf{MovieLens-1M (ML-1M)} \cite{he2017neural}. This dataset contains about one million movie ratings by about six thousand active users with over 20 ratings. It is a widely used benchmark for collaborative filtering algorithms. We convert it into implicit data by regarding rating actions as implicit feedback.
  \item \textbf{Pinterest} \cite{he2017neural}. This dataset contains user-image interactions (pins) in Pinterest. We use the version released from \cite{he2017neural} where they only retain users with more than 20 pins.
  \item \textbf{Yelp} \cite{shi2015semantic}. This dataset is originally released in the Yelp Challenge, which contains users' check-ins at different businesses, social networks and also cities and categories of the businesses. We filter out users with fewer than five check-ins as well as cold-start businesses for a fair comparison with the collaborative filtering baselines.
  \item \textbf{Douban} \cite{shi2015semantic}. Douban is a well-known social networking service in China. This dataset consists of users' movie ratings (ranging from 1 to 5), social relations and other metadata. Compared with ML-1M, the rating actions are highly biased towards popular movies, and thus we only regard ratings of five as positive feedbacks. We also filter out inactive users and cold-start items similarly as in Yelp.
\end{itemize}

The details of these datasets can be found in Table \ref{tab:dataset}.
We split each user's clicks into train (60\%), validation (20\%) and test (20\%). Note that we only leverage user behaviors in the training set to construct graphs for path-based models to prevent data leakage.

\subsubsection{Baseline Methods}
We compare our proposed approach with the collaborative recommendation methods (\textbf{ItemPop} based on item popularity, \textbf{BPR} \cite{rendle2009bpr}, \textbf{WMF} \cite{hu2008collaborative, pan2008one} and \textbf{VAE} \cite{liang2018variational}) and the hybrid models that exploit both content and collaborative signals for recommendation (\textbf{LibFM} \cite{rendle2012factorization}, heterogeneous graph embedding-based model \textbf{HERec} \cite{shi2018heterogeneous} and the base model in our framework \textbf{Proposed\textsubscript{base}} as introduced in Sect. \ref{prelim}). 

\begin{figure*}
	\includegraphics[width=17cm]{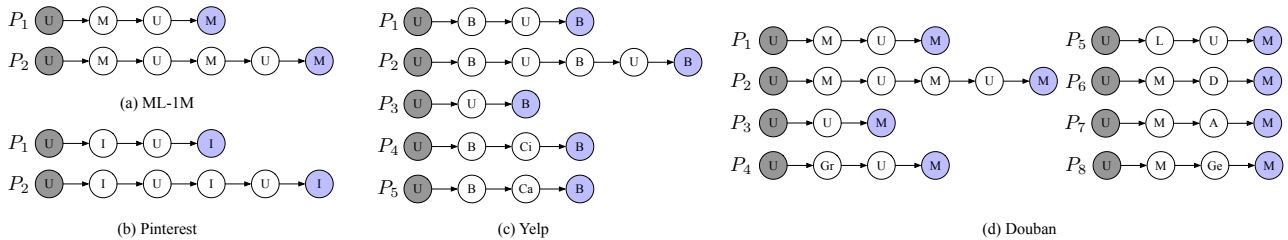}
  \centering
	\caption{Meta-paths for each dataset in our experiments. ($U$: User, $M$: Movie, $I$: Image, $B$: Business, $Ci$: City, $Ca$: Category, $Gr$: Group, $L$: Location, $D$: Director, $A$: Actor, $Ge$: Genre)}
	\label{fig:metapath}
\end{figure*}

\begin{table}[t]
  \centering
  \caption{Dataset Description.}
  \begin{adjustbox}{max width=\linewidth}
    \begin{tabular}{cccccc}
    \toprule
    Datasets & Relations (A-B)& \#A& \# B & \#A-B & Density \\
    \midrule
    ML-1M & \textbf{U}ser-\textbf{M}ovie & 6,040 & 3,706 & 1,000,209 & 4.468\% \\
    \midrule
    Pinterest & \textbf{U}ser-\textbf{I}mage & 55,187 & 9,916 & 1,500,809 & 0.274\% \\
    \midrule
    \multirow{4}[2]{*}{Yelp} & \textbf{U}ser-\textbf{B}usiness & 16,239 & 14,284 & 198,397 & 0.086\% \\
          & \textbf{U}ser-\textbf{U}ser & 16,239 & 16,239 & 158,590 & 0.060\% \\
          & \textbf{B}usiness-\textbf{Ci}ty & 14,284 & 47    & 14,267 & 2.125\% \\
          & \textbf{B}usiness-\textbf{Ca}tegory & 14,284 & 511   & 40,009 & 0.548\% \\
    \midrule
    \multirow{7}[2]{*}{Douban} & \textbf{U}ser-\textbf{M}ovies & 13,367 & 12,677 & 1,068,278 & 0.630\% \\
          & \textbf{U}ser-\textbf{U}ser & 13,367 & 13,367 & 4,085 & 0.002\% \\
          & \textbf{U}ser-\textbf{Gr}oup & 13,367 & 2,753 & 570,047 & 1.549\% \\
          & \textbf{U}ser-\textbf{L}ocation & 13,367 & 349   & 11,242 & 0.241\% \\
          & \textbf{M}ovie-\textbf{D}irector & 12,677 & 2,449 & 11,276 & 0.036\% \\
          & \textbf{M}ovie-\textbf{A}ctor & 12,677 & 6,311 & 33,572 & 0.042\% \\
          & \textbf{M}ovie-\textbf{Ge}nre & 12,677 & 38    & 27,668 & 5.744\% \\
    \bottomrule
    \end{tabular}%
  \end{adjustbox}
  \label{tab:dataset}%
\end{table}%

\subsubsection{Evaluation Metrics}
To evaluate the recommendation performance, we use user-item interactions in the training sets to predict items held out in the validation/test sets. For validation, held out items in the test set for each user are excluded from the candidate set and vice versa.
Three ranking metrics are used to evaluate the top-N recommendation results: Hit@K, Recall@K and NDCG@K. Slightly different from the one commonly used in literature, Recall@K is computed as in \cite{liang2018variational}, i.e., for each user $u$,
$$
Recall@K = \frac{\sum_{k=1}^{K}\mathbb I(\omega_u(k) \in I_u)}{\min(K, |I_u|)},
$$
where $\omega_u(k)$ is the $k$-th item recommended to user $u$ and the denominator is the minimum of $K$ and the number of relevant items $|I_u|$ (actually making itself a maximum of precision and recall). 
We use NDCG@K evaluated on the validation set to tune hyperparameters for the baselines and the proposed model, and then report the results achieved by the best configurations.

Meanwhile, it is hard to assess machine learning models' explainablility. With that being said, we can at least compare the faithfulness of model explanations given in the same form. Similar to the model fidelity metric defined in \cite{peake2018explanation}, we use the KL-divergence between the embedding-based model and the path-based model to measure the relevance of the explanations generated by the latter.

\subsection{Recommendation Performance}

To empirically investigate \textbf{RQ1}, we first evaluate the collaborative recommendation performance of our proposed model and baselines in ML-1M and Pinterest datasets. As shown in Table \ref{tab:reco_performance}, the proposed model significantly outperforms all the baselines in ML-1M and achieves the best results along with VAE in Pinterest. Compared with Proposed\textsubscript{base}, it improves NDCG@10 by 3.3\% and 11.4\% in ML-1M and Pinterest, respectively. This suggests that the proposed joint learning framework indeed helps embedding-based recommendation models to improve recommendation accuracy, which also partially answers \textbf{RQ3}.

In addition, we test the proposed approach in hybrid recommendation. Table \ref{tab:reco_meta_performance} shows the results in Yelp and Douban datasets where metadata are available. In Yelp, the proposed model achieves the best performance and the relative improvement in NDCG@100 over Proposed\textsubscript{base} is up to 14.0\%. In Douban, the recently proposed VAE outperforms the other methods, but our proposed model as the second best achieves very close results by improving its base model by 12.4\% in NDCG@100 and up to 15.5\% in Hit@5. 

Note that the results shown here only provide a lower bound for the performance of the proposed framework since we only use a simple base model (described in Sect. \ref{prelim}) for concise presentation.

\begin{table}[t]
\caption{Collaborative recommendation performance in ML-1M and Pinterest. The numbers in bold indicate statistically significant improvement ($p < .01$) by the pairwise t-test comparisons over the other baselines.}
  \centering
  \subtable[ML-1M]{\scalebox{1}{

    \begin{tabular}{rccc}
    \toprule
          & Hit@1 & Recall@10 & NDCG@10 \\
    \midrule
    ItemPop & 0.2520 & 0.1918 & 0.2039 \\
    BPR   & 0.4675 & 0.3629 & 0.3819 \\
    WMF  & 0.4803 & 0.3630 & 0.3837 \\
    VAE   & 0.4829 & 0.3707 & 0.3909 \\
    Proposed\textsubscript{base} & 0.4803 & 0.3630 & 0.3837 \\
    Proposed & \textbf{0.5005} & \textbf{0.3770} & \textbf{0.4002} \\
    \bottomrule

    \end{tabular}%
  \label{tab:ml1m}%
}  }
  \subtable[Pinterest]{  \scalebox{1}{

    \begin{tabular}{rccc}
    \toprule
          & Hit@5 & Recall@10 & NDCG@10 \\
    \midrule
    ItemPop & 0.0224 & 0.0072 & 0.0063 \\
    BPR   & 0.1592 & 0.0614 & 0.0514 \\
    WMF  & 0.1873 & 0.0705 & 0.0604 \\
    VAE   & 0.2335 & \textbf{0.0879}& \textbf{0.0774} \\
    Proposed\textsubscript{base} & 0.2168 & 0.0798 & 0.0685 \\
    Proposed & \textbf{0.2354} & \textbf{0.0872} & \textbf{0.0763} \\
    \bottomrule
    \end{tabular}%
  \label{tab:pinteret}%
}}
\label{tab:reco_performance}
\end{table}%

\begin{table}[t]
  \caption{Recommendation performance with metadata in Yelp and Douban. The numbers in bold and underlined indicate the best and the second best methods that are significantly better than the other baselines by the pairwise t-test at the level of $p < .01$, respectively.}
    \centering

  \subtable[Yelp]{\scalebox{1}{
    \begin{tabular}{rccc}
    \toprule
          & Hit@20 & Recall@20 & NDCG@100 \\
    \midrule
    ItemPop & 0.2000 & 0.0515 & 0.0593 \\
    BPR   & 0.1887 & 0.0478 & 0.0549 \\
    WMF   & 0.2755 & 0.0752 & 0.0859 \\
    VAE   & \underline{0.3293} & \underline{0.0911} & \underline{0.1058} \\
    LibFM & 0.2267 & 0.0582 & 0.0678 \\
    HERec & 0.2944 & 0.0788 & 0.0898 \\
    Proposed\textsubscript{base} & 0.3140 & 0.0863 & 0.0968 \\
    Proposed & \textbf{0.3476} & \textbf{0.1008} & \textbf{0.1104} \\
    \bottomrule
    \end{tabular}%
    }}
  \subtable[Douban]{  \scalebox{1}{
    \begin{tabular}{rccc}
    \toprule
          & Hit@5 & Recall@20 & NDCG@100 \\
    \midrule
    ItemPop & 0.3743 & 0.1920 & 0.2142 \\
    BPR   & 0.3485 & 0.1928 & 0.2063 \\
    WMF   & 0.3786 & 0.2079 & 0.2265 \\
    VAE   & \textbf{0.4403} & \textbf{0.2442} & \textbf{0.2645} \\
    LibFM & 0.3686 & 0.1978 & 0.2201 \\
    HERec & 0.3815 & 0.2261 & 0.2392 \\
    Proposed\textsubscript{base} & 0.3784 & 0.2077 & 0.2304 \\
    Proposed & \underline{0.4372} & \underline{0.2363} & \underline{0.2590} \\
    \bottomrule
    \end{tabular}%
    }}
\label{tab:reco_meta_performance}
\end{table}%
		
\subsection{Evaluation of the Joint Learning Process}
\label{eval_joint}
Although it has been shown that our approach achieves substantial improvement relative to its base model, the contribution of the proposed joint learning framework has not been thoroughly evaluated yet to answer \textbf{RQ3}. 
In Table \ref{tab:sparsity}, we further compare the recommendation performances of Proposed\textsubscript{base} and our proposed approach, from which we can draw the following two conclusions: 1) the proposed approach consistently improves its base model w.r.t. different data densities and 2) the joint learning process seems to be more effective when it comes to more sparse data.

We conduct an ablation study to demonstrate the effectiveness of joint learning in contrast with its alternatives. We compare with the following two variants of the proposed approach. In the first variant, namely \textit{Unlearnable}, we initialize all the parameters in the path-based model with the same values (i.e., uniform transition probabilities) and fix them at training time. In the second variant, \textit{Pipelined}, we first train the path-based model with $\mathcal L_2$ and then optimize $\mathcal L_1$ and $\mathcal L_3$ the same way as in Sect. \ref{training} but only with respect to the embedding model.

As shown in Table \ref{tab:ablation}, the unlearnable path-based model can even significantly improve the performance of the embedding-based model by 5.2\% and 5.5\% in Pinterest and Douban, respectively. This proves the effectiveness of label propagation via meta-paths. Then, the pipelined approach brings about another improvement in NDCG with the ability of ``learning to propagate.'' 

In Pinterest, the joint learning approach does not significantly outperform its pipelined variants in terms of recommendation accuracy with the embedding-based model. 
However, it helps the path-based model to obtain better predictive performance measured in NDCG and substantially minimizes its KL-divergence with the embedding-based model. These two measures can indirectly quantify the explainability of these approaches, indicating the \textit{quality} and the \textit{relevance} of the interpretations given by the path-based models, respectively.

In Douban, it is hard to train the path-based model separately, and it starts overfitting in even one iteration (hence, the performance of \textit{Pipelined} reported in Table \ref{tab:ablation}(b) is worse than that of \textit{Unlearnable}), although the overfitted model can still marginally boost the performance of the embedding-based model. 
On the contrary, the joint learning process can dramatically mitigate this issue and hence lead to large improvements on both sides over its pipelined variants. Besides, the KL-divergence is also reduced dramatically, meaning that our framework can find a proper path-based model to interpret the embedding-based model within the black-box.

\begin{figure}
	\includegraphics[width=8cm]{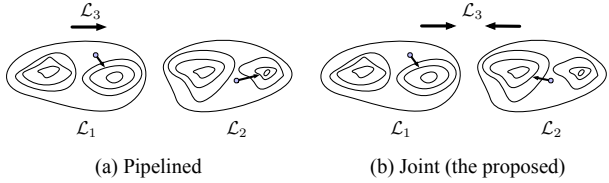}
	\caption{An intuitive comparison of the proposed joint learning approach with the pipelined alternative.}
	\label{fig:joint}
\end{figure}

Intuitively, we illustrate the difference between the joint learning approach and its pipelined alternative in Fig. \ref{fig:joint}. The problem at hand is by no means convex so there could exist many near-optimal local minima. The joint learning approach encourages the embedding-based model and the path-based model to find the closest local minima to each other simultaneously, while the pipelined approach can only regularize the embedding model.

Now we can draw a conclusion to \textbf{RQ3} that the proposed joint learning framework can orchestrate the embedding-based model and the path-based model for more accurate recommendation performance and better interpretability.

\begin{table}[t]
\caption{Recommendation performance in NDCG@10 and NDCG@100 w.r.t. different split ratios of the training sets in ML-1M and Yelp, respectively. \textbf{Improv.} denotes the performance improvement over Proposed\textsubscript{base}.}
\centering
\subtable[ML-1M]{
\scalebox{1.}{
\begin{tabular}{ccccc}
\toprule
\multicolumn{1}{c}{Ratio}& Density & Proposed\textsubscript{base} & Proposed & Improv. \\ \midrule
40\%& 1.787\%                & 0.3532  & \textbf{0.3713}   & +5.1\%  \\
60\%& 2.680\%                 & 0.3837  & \textbf{0.4002}   & +4.3\%  \\
80\%& 3.574\%                 & 0.4000  & \textbf{0.4162}   & +4.1\%  \\  \bottomrule

\end{tabular}}
}
\subtable[Yelp]{
\scalebox{1.}{
\begin{tabular}{ccccc}
\toprule
\multicolumn{1}{l}{Ratio} & Density & Proposed\textsubscript{base} & Proposed & Improv. \\ \midrule
40\%&0.034\%                 & 0.0827  & \textbf{0.0945}   & +14.2\% \\
60\%&0.051\%                & 0.0968  & \textbf{0.1104}   & +14.1\% \\
80\%&0.069\%                 & 0.1093  & \textbf{0.1203}   & +10.1\% \\ \bottomrule
\end{tabular}}
}
\label{tab:sparsity}
\end{table}

\begin{table}[t]
  \centering
  \caption{Ablation study. \textbf{Path} and \textbf{Embedding} represent recommendation performances in NDCG (NDCG@10 for Pinterest and NDCG@100 for Douban) of the path-based model and the embedding-based model in the joint learning framework, respectively. \textbf{Improv.} denotes the performance improvement of the embedding-based model over Proposed\textsubscript{base}. All the improvements are statistically significant at the level of $p<.01$.}
  \subtable[Pinterest]{
  \scalebox{1}{

    \begin{tabular}{rcccc}
    \toprule
          & Path  & Embedding & Improv. & KL-div. \\
    \midrule
    Proposed\textsubscript{base} & - & 0.0685 & +0.0\% & - \\
    Unlearnable & 0.0596 & 0.0720 & +5.1\% & 0.1569 \\
    Pipelined & 0.0702 & \textbf{0.0765} & +11.7\% & 0.3513 \\
    Proposed & \textbf{0.0740} & \textbf{0.0768} & +12.1\% & \textbf{0.1447} \\
    \bottomrule
    \end{tabular}%
  }}
  \subtable[Douban]{\scalebox{1}{
    \begin{tabular}{rcccc}
    \toprule
          & Path  & Embedding & Improv. & KL-div. \\
    \midrule
    Proposed\textsubscript{base} & - & 0.2304 & +0.0\% & - \\
    Unlearnable & 0.1980 & 0.2432 & +5.6\% & 0.7426 \\
    Pipelined & 0.1822 & 0.2458 & +6.7\% & 0.8101 \\
    Proposed & \textbf{0.2538} & \textbf{0.2590} & +12.4\% & \textbf{0.1892}\\
    \bottomrule
    \end{tabular}%
  }}
  \label{tab:ablation}%
\end{table}%

\subsection{Case Study}

\begin{figure}[t]
	\includegraphics[width=8.1cm]{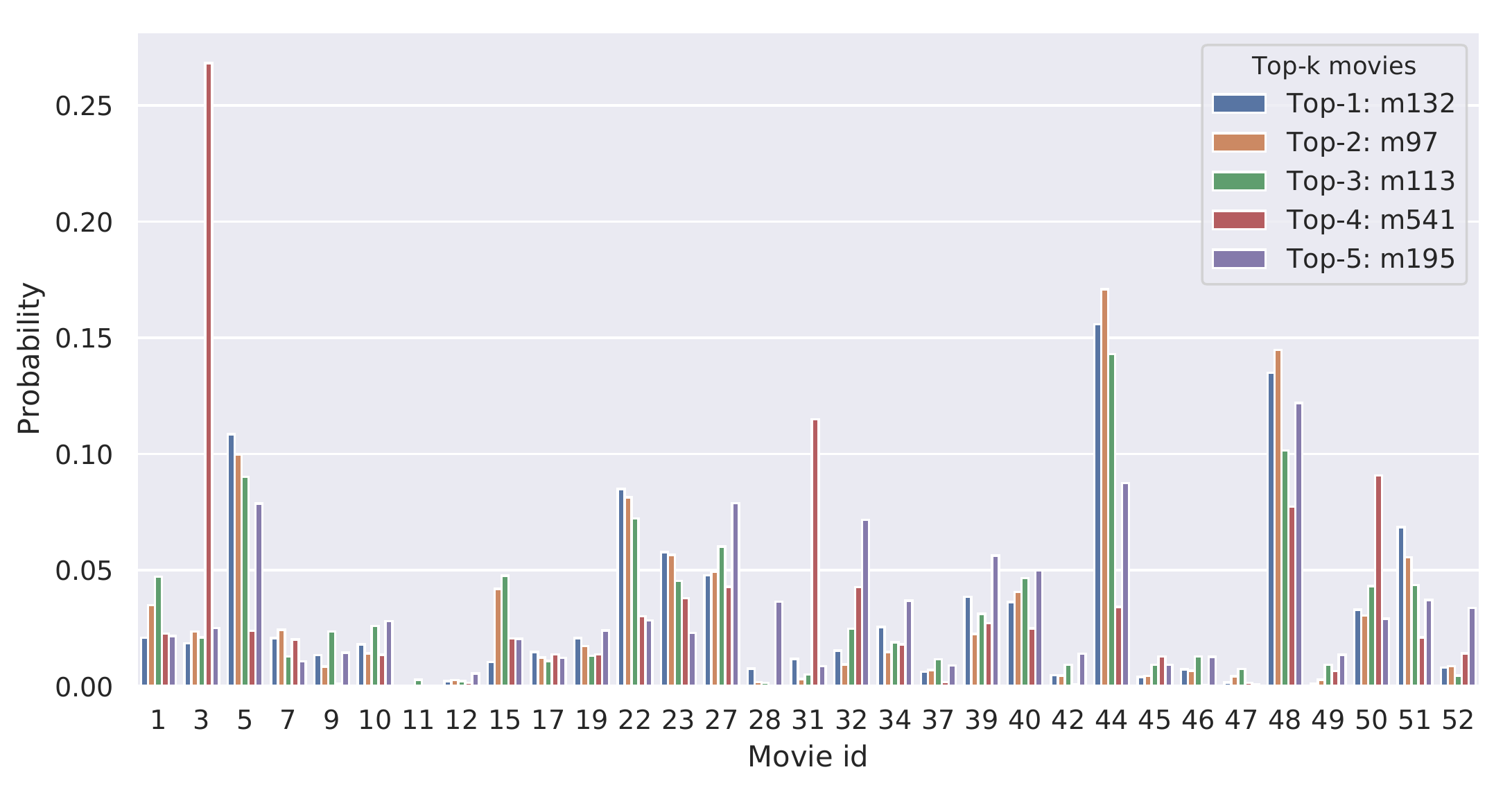}
	\caption{Relative contributions of watched movies to the top-5 recommendations to \textit{User 0} in ML-1M.}
	\label{fig:top5}
\end{figure}

\begin{figure}[t]
	\includegraphics[width=8.4cm]{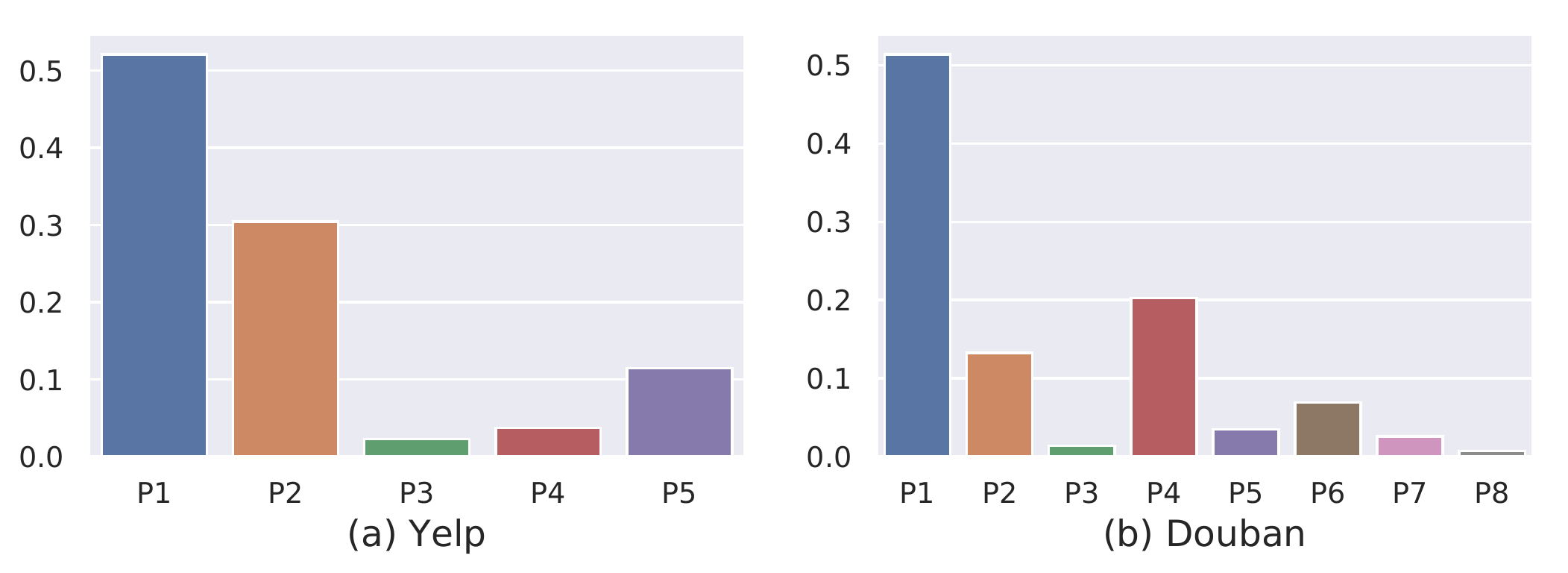}
	\caption{Average relative contributions of meta-paths in Top-20 recommendation in (a) Yelp and (b) Douban.}
	\label{fig:metapath_contri}
\end{figure}

To answer \textbf{RQ2}, we conduct case studies to demonstrate the interpretability of our approach. In Fig. \ref{fig:top5}, we show the relative importance of the movies that \textit{User 0} has watched with regard to the top-5 recommendations to her in ML-1M. 
From this explanation, we can draw the following observations. Some movies (e.g., \textit{movie 44} and \textit{48}) are consistently more informative than the others (especially \textit{movie 11} and \textit{12}). For different recommended items, they also contribute differently. For example, \textit{movie 22} and \textit{44} are the most important factors regarding the first recommended movie (\textit{movie 132}), while \textit{movie 3} and \textit{31} become the most dominating ones regarding the fourth recommendation (\textit{movie 541}). 

In the hybrid recommendation scenario, the interpretability is of more significance because there are more factors interleaved with each other in recommendation engines. The proposed approach can provide explanations at different levels. For instance, as Figure \ref{fig:metapath_contri} shows, we can calculate relative contributions of various meta-paths to interpret the importance of different types of information at the model level. Specifically, we can see the collaborative signals are the most dominating factor in both Yelp and Douban for $P_1$ and $P_2$ account for major contributions. The categories of business in Yelp (indicated by $P_5$) and user groups in Douban (indicated by $P_4$) also play relatively important roles.

Besides, as shown in in Table \ref{tab:case}, various factors have different contributions for different recommended movies; \textit{movie 737} is presented to \textit{user 22} mainly because of its director (indicated by $P_6$) and \textit{movie 10365} mainly because of collaborative signals as well as user groups (indicated by $P_1$ and $P_4$, respectively). 
Furthermore, Table \ref{tab:case} suggests that our approach can provide recommendation reasons at a more concrete level. We can see that \textit{movie 737} is recommended because \textit{user 22} has watched \textit{movie 6587} directed by the same director \textit{p558}; and \textit{movie 10365}, mainly because it is for fans of \textit{movie 6587} and \textit{movie 9609} the user has watched.

\begin{table}[t]
\centering
\caption{Case study: Recommendation reasons in Douban.}
\begin{adjustbox}{max width=\linewidth}
\begin{tabular}{@{}clc@{}}
\toprule
\multicolumn{1}{l}{\textbf{User 22}} & \multicolumn{2}{c}{watched movies: [m3866, m3870, m6587, m9609, m10895]} \\
\multicolumn{2}{l}{\textbf{Top2 Recommendation}} &  \\ \midrule
Movie id & \multicolumn{1}{c}{Recommendation reasons} & Prob. \\ \midrule
\multirow{4}{*}{m737} & R1. $P_6$: $User \rightarrow Movie \rightarrow Director \rightarrow Movie$ & 0.62 \\
 & \hspace{.2cm}   $u22 \rightarrow m6587 \rightarrow \textit{\textbf{p558}}  \rightarrow m737$ & \hspace{.1cm} 0.62 \\
 & R2. $P_1$: $User \rightarrow Movie \rightarrow User \rightarrow Movie$ & 0.19 \\
 & \multicolumn{2}{c}{...}  \\ \midrule
\multirow{6}{*}{m10365} & R1. $P_1$: $User \rightarrow Movie \rightarrow User \rightarrow Movie$ & 0.58 \\
 & \hspace{.2cm}   $u22 \rightarrow \textit{\textbf{m6587}} \rightarrow [u8895, u6566, ...] \rightarrow m10365$ & \hspace{.15cm}0.23 \\
 & \hspace{.2cm}   $u22 \rightarrow \textit{\textbf{m9609}} \rightarrow [u11380, u5391, ...] \rightarrow m10365$ & \hspace{.06cm} 0.11 \\
 & \hspace{.2cm}  ... & \hspace{.14cm} ... \\
 & R2. $P_4$: $User \rightarrow Group \rightarrow User \rightarrow Movie$ & 0.23  \\
 & \multicolumn{2}{c}{...} \\ \bottomrule
\end{tabular}
\end{adjustbox}
\label{tab:case}
\end{table}

\section{Related Work}
Our work is closely related to the following areas. 
We omit recent advances in deep recommender systems, which is orthogonal to our work, due to limited space and refer the reader to Zhang et al. \cite{zhang2017deep} for a more comprehensive review.

\textbf{Path-based Recommendation.}
The path-based recommendation has been widely studied in the literature. Yu et al. \cite{yu2014personalized} propose to use meta-paths \cite{sun2011pathsim} to diffuse user-item preferences and then exploit matrix factorization techniques to calculate latent vectors for users and items for implicit recommendation. Shi et al. \cite{shi2015semantic} extend this work to weighted paths for explicit recommendation. Catherine and Cohen \cite{catherine2016personalized} propose to use a logical reasoning system called ProPPR to integrate different meta-paths in a knowledge graph. Shi et al. \cite{shi2018heterogeneous} propose a heterogeneous network embedding method for recommendation (HERec) based on meta-path guided random walks. Zhao et al. \cite{zhao2017meta} propose to exploit matrix factorization to extract latent features from different meta-paths and then use factorization machines with Group lasso to fuse these features. Jiang et al. propose a learnable random walk model for more accurate recommendations. 
Wang et al. \cite{wang2018path} propose a path-constrained embedding approach for discriminating substitutable and complementary products.
Wang et al. \cite{wang2018ripple} recently propose an end-to-end framework RippleNet that leverages knowledge graph embeddings to propagate user preferences through paths in knowledge graphs with attention mechanism. 
Hu et al. \cite{hu2018leveraging} also utilizes attention mechanism to conduct the meta-path based recommendation.

\textbf{Interpretable Machine Learning.}
Interpretability has been a very hot topic in the machine learning community for a long time \cite{lakkaraju2016interpretable}. 
We briefly review some existing work that inspires us the most in this line of research.
Craven et al. \cite{craven1996extracting} propose to extract interpretable representations from neural networks with decision trees. LIME \cite{ribeiro2016should} proposed by Ribeiro et al. attempts to explain predictions of any given classifier by approximating its predictions locally with a sparse linear model that humans can understand. 
Meanwhile, Wu et al. \cite{wu2017beyond} propose to regularize deep models with decision trees at training time to improve their interpretability. Hu et al. \cite{hu2016harnessing} also propose to transfer structured information of logic rules into neural networks with posterior regularization techniques to reduce uninterpretability. 
Our work combines both directions of interpretation and regularization to enable explainable and accurate recommendation at the same time.

\textbf{Explainable Recommendation.} 
Explainable recommendation has also attracted a lot of attention in recent years \cite{zhang2018explainable}. Early approaches attempt to use topic models to generate intuitive explanations for recommendation results, e.g., \cite{wang2011collaborative, mcauley2013hidden, ling2014ratings}. Zhang et al. \cite{zhang2014explicit} propose EFM which aligns the latent dimensions with explicit product features for explainable recommendation.
TriRank \cite{he2015trirank} and SULM \cite{bauman2017aspect} utilize sentiment analysis techniques to extract aspects and user opinions to produce recommendation explanations. In addition to user reviews, Ren et al. \cite{ren2017social} incorporate social relations for better explanations. 
Catherine et al. \cite{catherine2017explainable} leverage knowledge graphs to generate recommendations together with their explanations with Personalized PageRank.
Ma et al. \cite{ma2019jointly} exploits induced rules from knowledge graphs for more explainablility.

We can see that most of the existing work attempts to provide post-hoc explanations for recommendation results with auxiliary information (reviews, social relations and images). On the contrary, our work takes a whole different perspective. We propose to interpret the given recommendation model per se with a comprehensible path-based model without introducing external information.

Recently, Peake and Wang \cite{peake2018explanation} propose a post-hoc explanation approach by training association rules on the output of a matrix factorization black-box model in a pipelined manner. Compared with their work, the proposed approach in this paper trains the black-box model jointly with the model used to interpret it. As discussed in Sect. \ref{eval_joint}, our treatment not only provides better interpretability than the pipelined alternative but also enables better recommendation performance of the black-box model. 

\section{Conclusion}
In this paper, we propose an end-to-end joint learning framework to combine both the advantages of embedding-based recommendation models and path-based recommendation models. Given an embedding-based model that produces black-box recommendations, the proposed approach can not only interpret its recommendation results but also regularize that model with structured information encoded in learnable paths for better performance. 
Extensive experimental studies in various public available datasets suggest that the proposed joint learning approach can substantially improve recommendation accuracy and achieve state-of-the-art performances. 
Through case studies, we also demonstrate that our approach can effectively provide intuitive explanations for recommendations made by black-box models at different levels.

\begin{acks}
This work is supported by National Key Research and Development Program of China under Grant No. 2018AAA0101902, NSFC under Grant No. 61532001, and MOE-ChinaMobile Program under Grant No. MCM20170503.
\end{acks}

\bibliographystyle{ACM-Reference-Format}
\bibliography{source}

\end{document}